\documentclass[prl,twocolumn,amssymb]{revtex4}
\usepackage{graphicx}

\begin{document}

\title{Critical Rotational Frequency for Superfluid Fermionic Gases across
Feshbach Resonance}
\author{Hui Zhai and Tin-Lun Ho}
\affiliation {Physics Department, The Ohio State University, Columbus, Ohio
43210}
\date{\today}
\begin{abstract}
We present a method to determine the critical rotational frequencies for superfluidity of both uniform and trapped Fermi gases across {\em wide} Feshbach resonance. It is found that as one approaches the resonance from the BCS side, beyond a critical scattering length, pairing is so robust that superfluidity cannot be destroyed by rotation. 
Moreover, the critical frequency has a sequence of jumps revealing the appearance of Landau levels, which are particularly prominent for systems up to a few thousand particles. 
For rotational frequency below an ``ultimate" critical frequency, defined to be the lowest frequency at which the center of the cloud goes normal,  a trapped gas has a superfluid core surrounded by a normal gas as seen in recent experiments[C. H. Schunck et.al. cond-mat/0607298]. 
\end{abstract}
\maketitle

The recent discovery of vortex lattice in a rotating Fermi gas of $^{6}$Li near Feshbach resonance by the MIT group\cite{ketterle,MITrecent} provides first direct evidence of phase coherence of this dilute system.  This is analogous  to the vortex lattice in type-II superconductors in the presence of magnetic field.  Like superconductors, which turn normal in sufficiently large magnetic fields, fermion superfluids will also turn normal for sufficiently large rotational frequency. 
This question of critical rotational frequency becomes even more  interesting at resonance since
fermion superfluidity is strongest at resonance and the properties of the system is universal 
in the absence of rotation\cite{Ho-uni}. As rotation increases, more angular momentum is deposited into the system, 
driving it toward the quantum Hall regime. How the fermion system copes with the strong pairing interaction and large angular momentum is an intriguing and fundamental question.

For harmonic traps,  the external rotation frequency $\Omega$ must be less than the radial frequency of the harmonic trap  $\omega$ in order to confine the system. The angular momentum of the system increases
as $\Omega$ increases, and becomes a maximum when $\Omega/\omega=1$. 
Since critical rotation frequency reflects the strength of superfluidity, it scales with fermion density and is therefore a decreasing function of radial distance, $\Omega_{c}= \Omega_{c}(r)$ . Thus, as $\Omega$ increases, the surface of the cloud will first turn normal, while the center region remains  superfluid with an array of vortices. At a particular frequency $\Omega_{c}^{\ast}$, the center region of the superfluid shrinks to zero and the whole cloud turns normal. It is this {\em ``utimate" critical frequency}
at very low temperatures that is the subject of our discussion. 

\begin{figure}[bp]
\begin{center}
\includegraphics[width=9.0cm]
{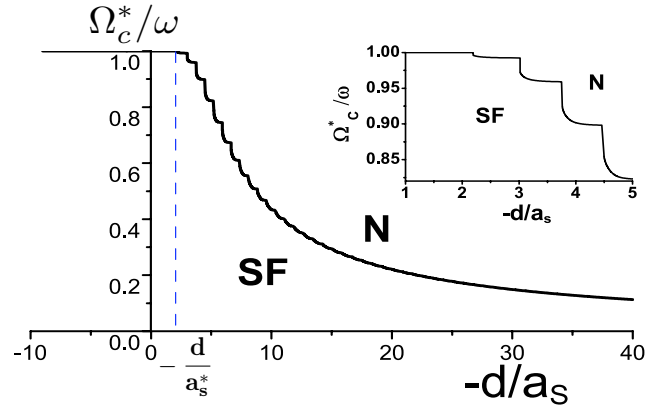}\caption{  
$\Omega^{\ast}_{c}$ is the ``ultimate" critical frequency defined as the critical frequency at the center of the trap.  
$a_{s}$ is scattering length. This diagram is for a system of 
$10^{3}$ fermions. 
The aspect ratio of the trap is  $\omega/\omega_{\text{z}}=2.5$, where 
$\omega$ and $\omega_{z}$ are radial and axial frequencies. $d=\sqrt{\hbar/m\omega}$.  The temperature  range is 
$k_{B}T/\hbar\omega = 10^{-3}$.  (SF) and (N) denote superfluid and normal phase respectively. 
The insert is a zoom-in plot of the strongly interacting regime. 
\label{phasediagram}}
\end{center}
\end{figure}

In this letter, we shall derive the phase diagram of the rotating
fermions at temperatures $k_{B}T<< \hbar \omega$ and discuss its unique features. The result is shown in fig.\ref{phasediagram}.   We find that: ${\bf (1)}$  As one approaches the resonance from the BCS side, $\Omega_{c}^{\ast}$ increases and has a sequence of steps which reflect the appearance of Landau levels. The size of these steps shrinks as particle number increases, but becomes sizable for systems with a few thousand particles.
(${\bf 2}$) $\Omega^{\ast}_{\text{c}}/\omega$ remains unity until the scattering length is beyond
a critical value $a_{s}^{\ast}$ that
depends on particle number, (see fig.\ref{phasediagram}).  This means the strong interaction near resonance is able to protect pairing even in the limit of large angular momentum. While this may not be obvious, it can understood by considering the opposite. If $\Omega^{\ast}_{c}< \omega$ at resonance, the system would be normal when $\Omega$ approaches $\omega$ and all  fermions would be in the lowest Landau level. As one makes the system more and more dilute, one eventually reaches a two-body problem. However, explicit solution of  two-body problem shows that at resonance, the fermions do not stay at the lowest Landau level, as higher Landau levels enable the fermions to come closer to take advantage of the strong interaction. To eliminate this contradiction, the ratio $\Omega^{\ast}_{c}/\omega$ can only be less than 1 on the BCS side of the resonance.

{\bf (A). How $\Omega^{\ast}_{c}$ is determined}: The Hamiltonian of a Fermi gas in a rotating harmonic trap with chemical potential $\mu$ is $K = H- \Omega L_{z} -\mu N = {\cal H}(\Omega)  - \int \mu({\bf r}) n({\bf r})$, where ${\cal H}(\Omega) = \sum_{i} h_{i}(\Omega)+ \sum_{i>j}U_{ij}$  is the many-body Hamiltonian of a uniform system with single particle Hamiltonian $h(\Omega)=({\bf p}-m\Omega\hat{{\bf z}}\times{\bf r})^2/(2m)$, and $U_{ij}$ is the interaction between particles.  The quantity $\mu({\bf r})$ is defined as $\mu({\bf r}) = \mu - V({\bf r})$, where  $V({\bf r})=m(\omega^2-\Omega^2){\bf r}_{\perp}^2/2+m\omega_{z}^2z^2/2$, ${\bf r}_{\perp}=(x,y)$. Our strategy is to first find $\Omega_{c}$ for the {\em uniform} system ${\cal K}(\Omega, \mu) = {\cal H} (\Omega) - \mu N$, and then use local density approximation to study the trapped case.
 
As we shall show later, the condition for superfluidity for a uniform system with chemical potential $\mu$ is that the pairing susceptibility $W$ (a function of $\mu/(\hbar\Omega)$ and defined later) to be sufficiently large, satisfying 
\begin{equation}
- a/a_{s} \leq W(\mu/\hbar\Omega), \,\,\,\,\,\, a = \sqrt{ \hbar/2m\Omega} = a(\Omega). 
\label{key} \end{equation}
The critical frequency $\Omega_{c}$ is determined by eq.(\ref{key}) with an equal sign.  
The number density of this uniform system is denoted as  $n$, and $n=n(\mu, \Omega)$. 

The situation in a trap is trickier, since in that case the chemical potential and density at ${\bf r}$ is 
$\mu({\bf r})= \mu - V({\bf r})$ and $n({\bf r}) = n(\mu({\bf r}), \Omega)$ respectively.  $\mu$ itself is determined by the number equation $N =\int n(\mu({\bf r}), \Omega)$, which gives
$N=N(\mu, \Omega)$. 
To find the critical frequency at ${\bf r}$, one inverts the number relation to get $\mu=\mu(N, \Omega)$, and then solves the equation  
$- a(\Omega)/a_{s} = W([\mu(N, \Omega)- V({\bf r})]/\hbar\Omega)$, which gives $\Omega_{c}({\bf r})$. 
In general, when the system has a superfluid core surrounded by a normal cloud,  the  equation
$N= \int n({\bf r})$ is quite complex for it requires the knowledge of $n$ of a {\em uniform} gas in both normal and superfluid phase. 
The situation is simplified considerably for the {\em ultimate} critical frequency $\Omega^{\ast}_{c}$. In that case, the whole cloud is normal, only the center is barely  superfluid. As a result, $n({\bf r})$ is simply the density $n_{N}^{}(\mu({\bf r}))$ of a normal gas in a rotating trap, where $n_{N}(\mu)$ is the number density of a {\em uniform} rotating normal gas, which is much easier to calculate; and $N=\int n_{N}^{}({\bf r}) = N_{N}^{}(\mu, \Omega)$. The corresponding expression for $\mu$ after inverting this equation is $\mu = \mu_{N}^{}(N, \Omega)$.  The equation for $\Omega^{\ast}_{c}$ is then 
\begin{equation} 
- \frac{d}{a_{s}} = \frac{d}{a(\Omega)} W\left(\frac{ \mu^{}_{N}(N, \Omega)}{\hbar\Omega}\right)\equiv G(N, \Omega), \,\,\,\,\,\,
\label{onset} \end{equation}
where $ d = \sqrt{\hbar/(m \omega)  }$.  The solutions of eq.(\ref{onset}) give  fig.\ref{phasediagram} .

{\bf (B1) Solving ${\cal H}(\Omega)$ in the Landau gauge}: For uniform systems, 
it is convenient to express the eigenstates of $h(\Omega)$ in terms of those in {\em  Landau gauge}: 
$\phi^{}_{\nu}(x,y,z) = e^{ikz} e^{im\Omega xy/\hbar} e^{-ipx}u_{np}(y)$,  where  ${\bf \nu}\equiv (n,p; k)$ labels the Landau level $n$ and degeneracy $p$,  and momentum $k$ along $z$; 
$u_{np}=H_{n}(y/a-pa)e^{-(y/a-pa)^2/2}/\sqrt{2^{n}n!\sqrt{\pi}a}$, $H_{n}$
is the Hermite polynomial with index $n$. We have applied periodic boundary condition along $z$ and $x$, taking the sample size $L_{z}$ and $L_{x}$ to be 1. 
We also find it is convenient to use two-channel model description. The eigenstates for close channel molecules $\Phi_{\nu}$ have the same form as $\phi_{\nu}^{}$, except that $m$ is replaced by $2m$ and
$a$ is replaced by $a/\sqrt{2}$.  It is also sufficient to write the two-channel model in the rotating frame as, 
\begin{equation}
{\mathcal H}(\Omega)=\sum\limits_{\nu\sigma}\xi_{\nu}f^\dag_{{\bf
\nu}\sigma}f_{{\bf \nu}\sigma} +\Xi_{0}b^\dag b+\alpha\sum\limits_{\nu\nu^\prime}
\left(Q_{\nu\nu^\prime}^*
b^\dag f_{\nu\uparrow} f_{\nu^\prime\downarrow}+\text{h.c.}\right)\label{Hamiltonian}, 
\end{equation}
where
$f^\dag_{\nu\sigma}$ creates an open channel fermion with quantum number $\nu$,  spin $\sigma$, and energy $\xi_{\nu}=\hbar^2k^2/(2m)+(2n+1)\hbar\Omega-\mu$. ($\xi_{\nu}$ is independent of $p$).  $b^\dag$ creates a close channel molecule. 
Since we shall focus on the ground state, and since the condensate occurs at the lowest energy, 
it is sufficient to restrict the Bose field $b$ to in the lowest Landau level and to zero momentum along $z$\cite{Ho}, with energy  $\Xi_{0}=\bar{\gamma}+\hbar\Omega-2\mu$, where $\bar{\gamma}$ is the unrenormalized detuning between close channel molecule and open channel fermions.  It is important to note that restricting the condensate to the lowest energy state imposes no restrictions on the quantum numbers of the fermions. 
Two fermions in {\em different} Landau levels can convert into a molecule in the lowest energy state, with the coupling
$\alpha$ and an overlap integral $Q^{}_{\nu\nu^\prime}=\int d{\bf
r}d{\bf R}\phi^*_{\nu}({\bf R}+{\bf r}/2)\phi^*_{\nu^\prime}({\bf R}-{\bf
r}/2)D({\bf R},{\bf r})$, where ${\bf R}$ and ${\bf r}$ are  the center-of-mass and relative coordinate of the Fermi pair, and 
$D({\bf R},{\bf r})$ is the  wavefunction of the tightly bound close-channel molecule, which can be taken as 
$\Phi({\bf R})\delta({\bf r})$. The field  $\Phi({\bf R})$ has zero momentum along $z$ and is  a {\em general  superposition} of in the lowest Landau level, i.e. 
$\Phi({\bf R})=\sum_{q}c_{q}\Phi_{(0,q,0)}$, where $c_{q}$'s are arbitrary except for the normalization 
$\sum_{q}|c_{q}|^2 =1$. An explicit calculation in Appendix I gives
\begin{equation}
Q^{}_{\nu\nu^\prime}=c^{}_{P}\frac{(-1)^{n}}{\sqrt{\sqrt{\pi}a}}\frac{H_{N}(\sqrt{2}\bar{p}a)e^{-(\bar{p}a)^2}}{2^{N}\sqrt{n!n^\prime!}}
 \delta_{k, - k'}^{}, 
\label{coefficient}
\end{equation}
where $\bar{p} = p-p^\prime$, $P=p+p^\prime$, and $N=n+n^\prime$. For later use, we also note that 
\begin{equation}
\sum_{p, P} |Q_{\nu\nu'}^{} |^2 =  \eta(n,N) \sum_{P}|c_{P}^{}|^2 =  \eta(n,N), 
\label{sumQ} 
\end{equation}
\begin{equation}
 \eta(n,N) =  \frac{1}{2^{3/2}_{}\pi  a^2 2^{N}}
\frac{ N!}{n!(N-n)! }. 
\label{eta} \end{equation}

{\bf (B2) Renormalzation}:  To solve the resonance model, it is essential to relate the parameters  ($\alpha$, $\bar{\gamma}$) to physical properties such as scattering length $a_{s}$ and effective range $r_{o}$. This is done by solving the two body problem. The procedure is explicitly shown in Appendix II. The result is 
that $\bar{\gamma}$ is formally divergent, and must be renormalized to a finite parameter as 
$\gamma$ $=$ $\bar{\gamma}$ $- { \alpha^2 m}/{(4\pi \sqrt{2}a\hbar^2)}\sum_{N}{1}/{\sqrt{N+1}}$.   The finite parameters $(\gamma, \alpha)$ must be chosen to produce the low energy scattering properties $(a_{s}, r_{o})$. The relation are 
\begin{equation}
-\frac{\gamma}{\alpha^2}=\frac{m}{a_{s}\hbar^2\pi\sqrt{2}} ;\   \ 
\frac{1}{\alpha^2}=-\frac{r_0 m^2}{2\pi \sqrt{2}\hbar^4}.\label{parameter}
\label{para}\end{equation}

{\bf (C1) Critical Frequency for uniform system}. For the many-body case, within mean-field
approximation $\alpha\langle {\bf b} \rangle=\Delta$,  ${\cal H}(\Omega)$ becomes
\begin{equation}
\mathcal{H}_{M}=\sum\limits_{\nu\sigma}\xi_{\nu}f^\dag_{\nu\sigma}f_{\nu\sigma}+\frac{\Xi_{0}\Delta^2}{\alpha^2}+\sum\limits_{\nu\nu^\prime}\left(\Delta^*Q^*_{\nu\nu^\prime}f_{\nu\uparrow}f_{\nu^\prime\downarrow}+\text{h.c.}\right).\label{mean-field-Hamil}
\end{equation}
Near $\Omega_{c}$, $\Delta$ is sufficiently small. The free energy
${\cal F}$ expanded in powers of $\Delta$ is
\begin{equation}
{\cal
F}=\frac{\Xi_{0}\Delta^2}{\alpha^2}-\sum\limits_{\nu\nu^\prime}\frac{\tanh(\frac{\beta\xi_{\nu}}{2})+\tanh(\frac{\beta\xi_{\nu^\prime}}{2})}{2(\xi_{\nu}+\xi_{\nu^\prime})}|Q_{\nu\nu^\prime}|^2\Delta^2+O(\Delta^4),
\end{equation}
where $\beta= 1/(k_{B}T)$ and the coefficient in front of $\Delta^4$ is
positive. 
Minimizing ${\cal F}$ with respect to $\Delta$, using the renormalization condition for $\bar{\gamma}$ and eq.(\ref{para}), and at the end using the condition for wide resonance $r_{o}/a \rightarrow 0$, we obtain the linearized gap equation, eq.(\ref{key}), 
\begin{eqnarray}
-\frac{a}{a_{\text{s}}}=&&\frac{\pi\sqrt{2}a\hbar^2}{2m}\sum\limits_{\nu\nu^\prime}\frac{\tanh(\frac{\beta\xi_{\nu}}{2})+\tanh(\frac{\beta\xi_{\nu^\prime}}{2})}{\xi_{\nu}+\xi_{\nu^\prime}}|Q_{\nu\nu^\prime}|^2\nonumber\\&&-\sum\limits_{N}\frac{1}{4\sqrt{N+1}}
\equiv  W\left(\frac{\mu}{\hbar\Omega}, \frac{k_{B}T}{\hbar\Omega}\right). 
\label{criticalEq}
\end{eqnarray}
Solution of Eq.(\ref{criticalEq}) gives the critical frequency $\Omega_{c}$ as a function of $\mu$ and temperature $T$.

Due to the constraint on $Q_{\nu\nu^\prime}$ in $W$, the sum $\sum_{\nu\nu'}$ is over $(p, P; n, n' ;  k)$, and 
$\xi_{\nu} + \xi_{\nu'}=  \hbar^2k^2/m + (2N+1) \hbar \Omega - 2\mu = E_{Nk} - 2\mu $.  If $\mu$ does not coincide with any Landau level, those Laudau levels below $\mu$ will lead a vanishing energy denominator for some {\em non-zero} value of $k$, which in turn causes the $k$-sum to be  logarithmic divergent if it were not  the $tanh$-factors in 
 eq.(\ref{criticalEq}) which eliminate this divergence at non-zero temperatures. On the other hand, if $\mu$ coincides with a Landua level $N$, the corresponding energy denominator vanishes at $k=0$, the $k$-sum becomes $\int {\rm d}k/k^2$ and is divergent. This will lead to a sequence of spikes in $W$ as a function of $\mu$.   

To express eq.(\ref{criticalEq}) in a form convenient for calculation at low temperatures, we note that $W$ is made up of  the following contributions:  (A) both $(\xi_{\nu}, \xi_{\nu'})$ have the same sign, (B) both $(\xi_{\nu}, \xi_{\nu'})$ have opposite signs, corresponding to fermions on the same side or opposite sides of the chemical potential respectively.  Contribution (A) can  be further divided into: (A1) $|\xi_{\nu}|,  |\xi_{\nu'}|>k_{B}T$, (A2) $|\xi_{\nu}|,  |\xi_{\nu'}|<k_{B}T$.  In the case of (A1), the $tanh$-terms in eq.(\ref{criticalEq}) can be replaced by 1.  On the other hand, the $tanh$-terms make the summand of (A2) and (B) finite, and therefore  become insignificant as $T\rightarrow 0$. One can therefore approximate $W$ for $k_{B}T<< \hbar\Omega$ as 
\begin{eqnarray}
W\left(\frac{\mu}{\hbar\Omega}, \frac{k_{B}T}{\hbar\Omega}\right) = 
\frac{\pi\sqrt{2}a\hbar^2}{2m}
\sum\limits_{\nu \nu'}  \frac{[ \epsilon(\xi_{\nu}) + \epsilon(\xi_{\nu'})] |Q_{\nu\nu'}|^2}
{ [ (\xi_{\nu} + \xi_{\nu'})^2 + (k_{B}T)^2]^{1/2}} \nonumber \\
    -\sum\limits_{N}\frac{1}{4\sqrt{N+1}}, \hspace{1.0in}
\label{WW} \end{eqnarray}
where $\epsilon(x) = \pm 1$ if $x  ^{>}_{<} 0$.
Note that both $\xi_{\nu}$ and $\xi_{\nu^\prime}$ are independent of
$p$ and $p^\prime$, we can therefore sum over the degenerate index $P$
and $\bar{p}$ in eq.(\ref{WW}) as in the two atom case and obtain a result 
independent of  $\{c_{P}\}$. This shows $\Omega_{c}$ is insensitive to the
configurations of vortices at their onset in the superfluid phase, as in type-II superconductor. 
We then obtain
\begin{equation}
W =  \sum\limits_{N}\left(\frac{a \hbar^2}{2^{3/2} m}   \int_{0}^{\infty} 
\frac{ {\rm d}k  {\cal L}(N,k)}{ [ (E_{Nk} - 2\mu)^2 + \delta^2 ]^{1/2} } - \frac{1}{4\sqrt{N+1}} \right). 
\label{W4} \end{equation}
\begin{equation}
{\cal L}(n,k) = \sum_{n=0}^{N}  [\epsilon(\xi_{\nu}) + \epsilon(\xi_{\nu'}) ] \eta_{n,N}^{}, 
\label{L} \end{equation}
where $\nu = (n, p; k )$ and $\nu' = (N-n, p', -k)$.  While many more simplification of can be made, 
which we shall not present for length reasons, eq.(\ref{W4}) and (\ref{L}) are in a form convenient for numerical calculations.  We have calculated $W$  for a system of $10^{3}$ fermions from eq.(\ref{W4}) for $k_{B}T ^{<}_{\sim} 10^{-3}\hbar\Omega$ and have verified numerically that $W$ changes insignificantly  even when $T$ is lowered by another factor of 10. Thus, from now on, we shall not display the $T$ dependence of $W$ explicitly. 

\begin{figure}[tbp]
\begin{center}
\includegraphics[width=8.7cm]
{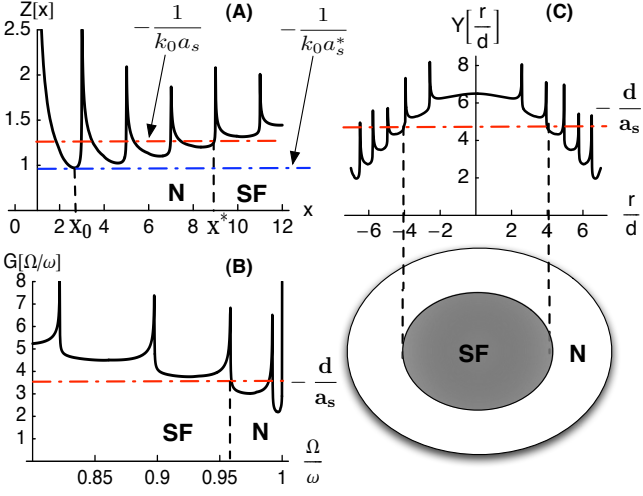}\caption{(A)$Z(x)$ for a uniform system,
$x=\mu/(\hbar\Omega)$. As one approaches resonance from the BCS side, the horizontal line comes down. The last intersection takes place at scattering $a_{s}^{\ast}$ when the horizontal line touches the minimum of $Z(x)$.  (B) The function $G(\Omega/\omega)$
plotted for $N_{0}=10^3$ fermions 
and the aspect radio $\omega/\omega_{z}=2.5$. (C) The function of
$Y(r/d)$  plotted for $\mu=10\hbar\omega$,  $\omega/\omega_{z}=2.5$, $\Omega/\omega=0.8$ and
$z=0$.\label{spatial}}
\end{center}
\end{figure}

To determine the critical frequency $\Omega_{c}$, we rewrite  eq.(\ref{criticalEq}) as $- 1/(k_{o} a_{s}) = x^{-1/2} W(x)\equiv Z(x) $, where $\hbar^2 k_{o}^{2}/(2m) = \mu$, $x= \mu/(\hbar \Omega)$.
Fig.\ref{spatial}(A) displays the solutions of this equation,  which are the interactions between $Z(x)$ and the horizontal line  $y= -1/(k_{o}a_{\text{s}})$. The spikes in $Z(x)$ is due to the presence of Landau levels. 
The regions of $x$ where $Z> -1/(k_{o} a_{\text{s}})$ and $Z< -1/(k_{o} a_{\text{s}})$ correspond to superfluid and normal phase respectively. The largest intersection $x^{\ast}$ corresponds to the lowest rotational frequency at which the superfluid turns normal. Hence, we identify $\Omega_{c} = (\mu/\hbar)/x^{\ast}$.  Since $Z(x)$ is positive definite,  eq.(\ref{criticalEq}) only has a solution if $a_{s}<0$, i.e. on the BCS side of the resonance. Moreover, since $Z$ has a minimum (located at $x=x_{o}=2.67$),  eq.(\ref{criticalEq})  has no solution for $k_{o}a_{s}> k_{o}a_{s}^{\ast} \equiv - 1/Z(x_{o})$, $Z(x_{o})= 0.9716$.  This means approaching the resonance from the BCS side, the critical frequency will rise to $\Omega_{o} = (\mu/\hbar)/x_{o}$  at $k_{o}a_{s}^{\ast}=  -1/ Z(x_{o})= -1.029$.  
For scattering length  $a_{s}> a_{s}^{\ast}$, the system will remain supefluid for all $\Omega$.

{\textbf{(C2) ``Ultimate" critical frequency $\Omega_{c}^{\ast}$ in a trap}}.  
The scheme of finding 
$\Omega_{c}^{\ast}$ has been discussed in {\bf (A)}. We shall start with the number density of a 
{\em uniform} rotating normal system, which is 
$n_{N}(\mu, \Omega) = 2\sum_{npk} \Theta(\mu - (2n+1)\hbar\Omega - \hbar^2k^2/(2m))$, where 
$\Theta(x)=1$(0) if $x>0 (<0)$.  Summing over $k$ and using local density approximation, we have
\begin{equation}
n_{N}^{}(\mu({\bf r}), \Omega) = \frac{1}{\pi^2 a^3} \sum^{n^{\ast}}_{n=0} \sqrt{
 \frac{\mu({\bf r})}{\hbar \Omega} - 2n-1   }, 
\label{nN} \end{equation}
where $n^{\ast} = [  \mu({\bf r})/(2\hbar\Omega) -1/2]$ and $[x]$ means the largest integer smaller than $x$. Eq.(\ref{nN}) allows one to find the relation between $\mu$ and $N$ in the rotating normal state, and to construct the function $G(N, \Omega)$. 
In Fig.\ref{spatial}(B), we have shown the function  $G(N, \Omega)$ for an anisotropic trap $(\omega/ \omega_{z}=2.5)$ (like that in ref.\cite{ketterle}) with the scaled variable 
 $\Omega/\omega$.  The {\em ultimate} critical frequency $\Omega^{\ast}_{c}$  is given by the intersection between $G(\Omega/\omega)$ and the horizontal line $-d/a_{s}$ with smallest $\Omega$. It is from these intersections that fig.\ref{phasediagram} is constructed.   
It should be noted that $\Omega/\omega$ can at most be 1, and $G(\Omega/\omega)$ is positive definite with a global minimum at $\Omega^{\ast}/\omega$ close to 1. The situation is almost identical to ${\bf (C1)}$: $\Omega^{\ast}_{c}$ exists only on the BCS side of the resonance. It undergoes a series of jumps as one approaches the resonance, reflecting the presence of Landau levels. It will finally reaches $\Omega^{\ast}$ when the scattering length reaches a critical value $a^{\ast}_{s}$ such 
that $-d/a_{s}^{\ast} = G(\Omega^{\ast}/\omega)$.  For all $a_{s}> a^{\ast}_{s}$, the system remains a superfluid even when $\Omega/\omega$ reaches 1.  
Calculations with larger number of atoms show that the phase boundary in fig.\ref{phasediagram} will be pushed up to
higher rotating frequency and the size of Landau step will shrink.  However, the basic features will still hold. For systems with few thousand particles, (see fig.1), the Landau steps are highly visible.

{\bf (C3) Separation of vortex and normal regions:} Eq.(\ref{key}) also provides a convenient means to determine the spatial distribution between vortex and normal regions as found in the MIT experiment\cite{MITrecent}, as the superfluid region must have a sufficiently large pairing susceptibility.  Rewriting eq.(\ref{key}) in dimensionless form, and in local density approximation, condition for superfluidity at ${\bf r}$ is 
$-{d}/{a_{s}} \leq  \sqrt{{2\Omega}/{\omega}} W\left( {\mu - V({\bf r})}/{(\hbar \Omega)}\right) = Y({\bf r}/d)$.
In fig.2c, we have plotted $Y({\bf r})$ as a function of ${\bf r}$. The region of superfluid can be identified easily from the intersection between $Y$ and the horizontal line $y = -d/a_{s}$\cite{rings}.

We have thus established the results ${\bf (1)}$ and ${\bf (2)}$ mentioned in the beginning. Finally we would like to point out that although we consider the case where $k_{B}T/(\hbar\Omega)\sim10^{-3}$, our results should hold as long as $k_{B}T/(\hbar\Omega)<1$. This condition is satisfied in traps 
with frequencies about $300$ Hz ($\sim10^{-8}$K) at the lowest temperature attainable today ($10^{-9}$K).  We would also like to point out that although mean field theory is not exact, it is known that it provides a good qualitative description at very low temperatures. Fluctuations effects at temperatures higher than $\hbar \Omega$  will be considered elsewehre.

{\bf  Appendix I}. The coefficient $Q_{\nu\nu^\prime}$ is given by $\sum_{q}c_{q}\int d^3{\bf R}\phi^*_{\nu}({\bf R})\phi^*_{\nu^\prime}({\bf  R})\Phi_{q}({\bf R})$, which is 
$Q_{\nu\nu^\prime}=\sum_{q}c_{q}{I_{\nu\nu^\prime}\delta(p+p^\prime-q)}/{\sqrt{\pi^{3/2}a2^{(n+n^\prime-1)}n!n^\prime!}}$,
denoted as (${\ast}$),
where $
I_{\nu\nu^\prime}=\int_{-\infty}^{+\infty}dt H_{n}(t-t_0)H_{n^\prime}(t+t_{0})e^{-t^2}e^{-(t^2+t_0^2)} $,
 $t_0=(p-p^\prime)a$, and $t=y/a$. Using the well known relations $H_{n}(t)=(-1)^ne^{t^2}d^{n}(e^{-t^2})/dt^n$ and   $e^{-s^2+2(t-t_0)s}=\sum_{n=0}^{+\infty}H_n(t-t_0)s^{n}/n!$, integration by parts and straightforward calculations gives 
$
 I_{\nu\nu^\prime}=(-1)^{n}\sqrt{{\pi}/{2^{(n+n^\prime+1)}}}H_{n+n^\prime}(\sqrt{2}t_{0})e^{-t_0^2}.
$
Substituting this result back to eq.($\ast$) gives eq.(\ref{coefficient}).

{\bf  Appendix II}.  The two-body eigenstate with energy $E$ is  $|\Psi\rangle = [ B b^{\dagger} + \sum_{\nu\nu'}  A^{}_{\nu \nu'}  f^{\dagger}_{\nu \uparrow}f^{\dagger}_{\nu,  \downarrow}]|0\rangle$, where $(E-\overline{\gamma})B = \alpha \sum_{\nu\nu'} Q_{\nu\nu'}A_{\nu\nu'}$,  $A_{\nu\nu'}(k) = \alpha Q^{\ast}_{\nu\nu'} B/(E - E_{Nk})$,  where $E_{Nk}=2(N+1)\hbar\Omega+\hbar^2k^2/m$. Hence, we have  $E-\bar{\gamma}=\alpha^2\sum_{\nu\nu^\prime}{|Q_{\nu\nu^\prime}|^2}/{(E-E_{Nk})}$. 
Eq.(\ref{sumQ})-(\ref{eta}) imply $\sum_{n, P,\bar{p}} 
|Q_{\nu\nu^{\prime}}|^2 = \sum_{n=0}^{N} \eta(n,N) =  1/(2^{3/2}\pi a^2)$. 
The equation for bound states is 
 $\bar{\gamma}-E ={\alpha^2 m}/{(4\pi\sqrt{2}a\hbar^2)}\sum_{N}{1}/{\sqrt{N+1-E/(\hbar\Omega)}}$. 
The divergence in the sum 
can be removed using the renormalized detuning $\gamma$ defined in section ({\bf B2}), and we have $\gamma-E={\alpha^2 m}/{(4\pi\sqrt{2}a\hbar^2)}\sum_{N}({1}/{\sqrt{N+1-E/(\hbar\Omega)}}-{1}/{\sqrt{N+1}})$.
To relate  $(\gamma, \alpha)$ to $(a_{s}, r_{0})$, we take the limit $\hbar\Omega\rightarrow 0$, which yields $\gamma-E=-\alpha^2 m^{3/2}\sqrt{-E}/(\pi\sqrt{2}\hbar^3)$.  Comparing it to equation derived from scattering theory, $\sqrt{-E}=\hbar/(a_{s}\sqrt{m})-r_{0}\sqrt{m}E/(2\hbar)$, we find eq.(\ref{para}). 

This work is supported by NSF Grant DMR-0426149 and PHY-05555576.


\begin{thebibliography}{99}
\bibitem{rings}
Fig \ref{spatial}(C) also indicates that there are shells of superfluid
in the normal regime, where the local chemical potential happens to be
$(2N+1)\hbar\Omega$. However, the width of the shells is so narrow (much smaller than harmonic length) that the local density approximation will breakdown.
\end{thebibliography}
\end{document}